\documentclass[10pt,twocolumn,letterpaper]{article}
\usepackage{authblk}
\usepackage{cvpr}
\usepackage{times}
\usepackage{epsfig}
\usepackage{graphicx}
\usepackage{amsmath}
\usepackage{amssymb}

\usepackage{diagbox}
\usepackage{array}
\usepackage{makecell}
\usepackage{multirow}
\usepackage{multicol}
\usepackage{float}

\begin{document}

\title{BehaveFormer: A Framework with Spatio-Temporal Dual Attention Transformers for IMU enhanced Keystroke Dynamics}



\author[1]{Dilshan Senerath}
\author[1]{Sanuja Tharinda}
\author[1]{Maduka Vishwajith}
\author[2]{Sanka Rasnayaka}
\author[1]{Sandareka Wickramanayake}
\author[1]{Dulani Meedeniya}
\affil[1]{University of Moratuwa}
\affil[2]{National University of Singapore} 
\affil[ ]{\{dilshan.18, sanuja.18, maduka.18, sandarekaw, dulanim\}@cse.mrt.ac.lk, sanka@nus.edu.sg }

\maketitle

\thispagestyle{empty}

\begin{abstract}
Continuous Authentication (CA) using behavioural biometrics is a type of biometric identification that recognizes individuals based on their unique behavioural characteristics, like their typing style. However, the existing systems that use keystroke or touch stroke data have limited accuracy and reliability. To improve this, smartphones' Inertial Measurement Unit (IMU) sensors, which include accelerometers, gyroscopes, and magnetometers, can be used to gather data on users' behavioural patterns, such as how they hold their phones. Combining this IMU data with keystroke data can enhance the accuracy of behavioural biometrics-based CA. This paper proposes BehaveFormer, a new framework that employs keystroke and IMU data to create a reliable and accurate behavioural biometric CA system. It includes two Spatio-Temporal Dual Attention Transformer (STDAT), a novel transformer we introduce to extract more discriminative features from keystroke dynamics. Experimental results on three publicly available datasets (Aalto DB, HMOG DB, and HuMIdb) demonstrate that BehaveFormer outperforms the state-of-the-art behavioural biometric-based CA systems. For instance, on the HuMIdb dataset, BehaveFormer achieved an EER of 2.95\%. Additionally, the proposed STDAT has been shown to improve the BehaveFormer system even when only keystroke data is used. For example, on the Aalto DB dataset, BehaveFormer achieved an EER of 1.80\%. These results demonstrate the effectiveness of the proposed STDAT and the incorporation of IMU data for behavioural biometric authentication. The code is available at \url{https://github.com/DilshanSenarath/BehaveFormer}.

\end{abstract}

\section{Introduction}

The rapid development in the mobile phone industry has enabled people to easily carry out vital applications through their smartphones. These include applications regarding communication, finance, health, transportation and many more. Most of these mobile applications require access to sensitive user data. Since mobile devices could easily be shared or stolen, these sensitive data may get exposed easily. Thus, the requirement for robust security has made authentication a major area of importance in the mobile phone industry. 

Existing mobile authentication methods can be categorized into two main categories which are knowledge-based and physiological biometrics-based. Knowledge-based authentication methods include pin codes, passwords and pattern unlocks. Physiological biometrics includes fingerprints, face and iris recognition. However, both these authentication methods have security vulnerabilities. Guessing, sniffing, social engineering attacks, and shoulder surfing attacks are vulnerabilities of knowledge-based authentication methods \cite{kbattacks}. Presentation attacks, replay attacks and data simulation are vulnerabilities that can occur with physiological biometrics \cite{1495927}. Both these categories of authentication are one-time/session-based authentication systems. Once authenticated, there is no authentication performed until the current session is terminated. Furthermore, both these categories of authentication systems require the users' active participation to carry out a specific authentication task which hinders the usability \cite{8698599}. CA aims to address these issues.

CA focuses on continuously authenticating the user while the user is using the device. Since the verification is done throughout the session, the issues caused by one-time authentication systems no longer occur. Furthermore, CA does not require the active participation of the user, increasing usability. CA is achieved passively through behavioural biometrics. Behavioural biometrics can be defined as the use of unique behavioural traits of the user such as typing, use of touch gestures, and gait data to identify the user. Among these, the most commonly used and popular behavioural biometric is Keystroke Dynamics, which is the analysis of the user’s typing behaviour for identification. Furthermore, the latest trends focus on using IMU data in addition to keystroke dynamics or touch stroke dynamics to increase accuracy.

Keystroke analysis can be categorized into two types; (1) fixed text and (2) free text analysis. In fixed text analysis, the same text that was used in the enrollment phase must be used in the verification phase. In free text analysis, the text that is used in the verification phase need not be the same text which was used in the enrollment phase.  Free text analysis is considered to be more challenging than fixed text analysis. 

Most smartphones nowadays consist of IMU sensors, since they are affordable and useful. IMU data includes accelerometer data, gyroscope data and magnetometer data. Using IMU data, characteristics such as how the device is held and tilted while the user is using the device can be captured. This auxiliary information can provide a means to increase the discriminative power of biometrics such as keystroke and touch stroke.

This study proposes BehaveFormer, a new behavioural biometric-based authentication framework that utilizes keystroke and IMU data. The framework includes two Spatio-Temporal Dual Attention Transformers (STDAT), one for keystroke data and another for IMU data. STDAT is a novel transformer we introduce to extract more discriminative features from keystroke dynamics. It employs a dual attention module that concentrates on time and channel dimensions of keystroke dynamics separately. To evaluate the BehaveFormer, both traditional (EER) and continuous (Usability, TCR - Time to Correct Reject, FRWI - False Reject Worse Interval, FAWI - False Accept Worse Interval) evaluation metrics are used~\cite{10.1109/TPAMI.2007.1010,10.1109/TC.2016.2622262,10145743}. Our experiments with three public datasets (Aalto DB, HMOG DB, and HuMIdb) have shown that BehaveFormer outperforms the current state-of-the-art for keystroke dynamics in various evaluation settings. 
Additionally, we demonstrate the potential of using Transfer Learning with BehaveFormer to tackle dataset sizes in behavioural biometrics.

\section{Related Work}

Behavioural biometrics-based mobile CA systems have primarily been carried out with keystroke dynamics. Initially, models were built with traditional ML classifiers~\cite{electronics10141622,10.1145/3230820.3230829}. These methods required a significant effort in the feature extraction process. With the emergence of deep learning approaches and the popularity they had gained over other fields, they have also been employed in keystroke dynamics based CA systems as well. These deep learning approaches include Multi-Layer Perceptron, Recurrent Neural Networks, Long Short Term Memory (LSTM), and the latest trend includes Transformers. Deep learning approaches typically require a high amount of data to provide accurate results~\cite{zhang2021understanding}. The availability of large public datasets such as AaltoDB \cite{10.1145/3338286.3340120} alleviated this challenge. Other than keystroke dynamics, recent studies include IMU data to further improve the discriminative powers of behavioural biometrics. Rather than IMU data being used as a single measure for user authentication, recent studies have focused on multimodal architectures where different types of IMU data have been used along with other more discriminative behavioural biometrics which are primarily keystroke dynamics and touch dynamics. 

In \cite{electronics10141622}, De-Marcos \etal compared seven ML classifiers for keystroke dynamics based CA. Their investigation concluded that it was feasible to use a smaller number of key events and measurements for user identity prediction. Ensemble methods yielded the highest results. In \cite{Acien2022}, TypeNet, a LSTM based keystroke dynamics CA system was introduced. The training process for TypeNet was carried out with three loss functions, with triplet loss yielding the best results. TypeNet’s results obtaining was done for both identification and authentication. In \cite{stragapede_mobile_2022}, TypeFormer, a transformer based keystroke dynamics CA system was introduced yielding the best keystroke dynamics based EER and it was further improved in \cite{stragapede2022typeformer} (TypeFormer-BR). 

In \cite{Stragapede2022,stragapede2022mobile,https://doi.org/10.48550/arxiv.2206.02502}, Stragapede \etal introduced multimodal architectures for behavioural biometrics based CA. A separate LSTM model was developed and trained for each modality considered, and the final result was obtained with a score-level fusion. The different modalities were based on keystrokes, touch tasks and IMU data. The experiment was carried out with different combinations of modalities. For \cite{Stragapede2022} (HuMINet), the best results were achieved with keystroke, accelerometer and magnetometer combination. In \cite{https://doi.org/10.48550/arxiv.2206.02502}, keystroke, linear accelerometer, gyroscope, and magnetometer combination gave the best results. In \cite{stragapede2022mobile} (DuoNet), the drawing of “8” touch task, accelerometer, gyroscope and magnetometer combination gave the best results. In each of these works, the effect on a primary behavioural biometric, by the combination of IMU data, especially accelerometer, gyroscope, and magnetometer was highlighted. However, no study has considered transformers for a multimodal architecture for behavioural biometrics-based CA.  Furthermore, we use feature-level fusion rather than score-level fusion. Finally, none of these works evaluated their systems with continuous evaluation metrics~\cite{10.1109/TPAMI.2007.1010,10.1109/TC.2016.2622262}.

\begin{figure*}
\begin{center}
   \includegraphics[width=0.9\linewidth]{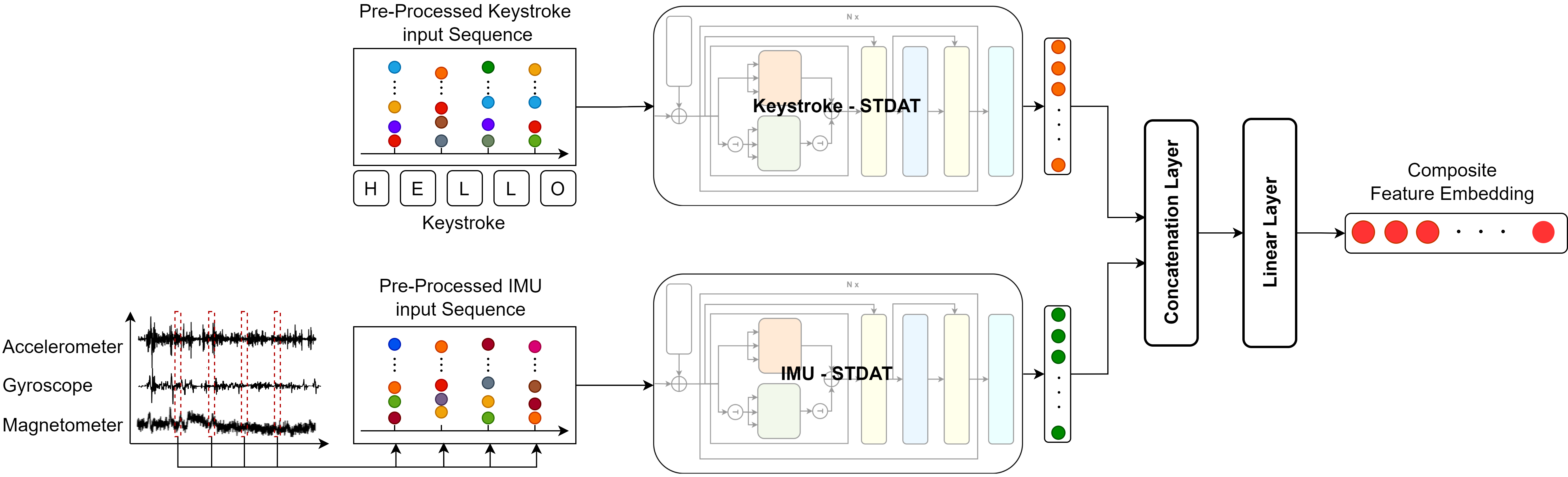}
\end{center}
   \caption{The overview of BehaveFormer. The top STDAT processes keystroke data, and the bottom STDAT processes IMU data. The feature embeddings from both STDATs are concatenated to create a composite feature embedding.}
   \label{fig:behaveformer}
\end{figure*}

\section{BehaveFormer Framework}

The BehaveFormer framework proposed in this paper uses Keystroke and IMU data (accelerometer, magnetometer, and gyroscope data) to improve the accuracy of behavioural biometric-based CA. To achieve this, we first preprocess and extract features from Keystroke and IMU data and then pass them to BehaveFormer. Next, we use a novel transformer called Spatio-Temporal Dual Attention Transformer (STDAT) to extract more discriminative features from keystroke dynamics. BehaveFormer employs two STDATs, one for each keystroke dynamic. Finally, we combine the features from both STDATs and process them through a linear layer to generate the final composite feature embedding. The overview of the proposed BehaveFormer is shown in Figure ~\ref{fig:behaveformer}. 

We design the training objective of BehaveFormer, $L$ as a Triplet Loss. $L$ helps BehaveFormer to learn an embedding function that maps the keystroke dynamics input to a vector space such that similar data points (input sequences from the same user) are mapped to nearby points in the vector space, while different data points (input sequences from different users) are mapped to far away points. Let BehaveFormer's output feature embedding for two input sequences of user $u$ be $f^u_1$ (Anchor) and $f^u_2$ (Positive), respectively, and $f_v$ (Negative) for the input sequence of user $v$. Then $L$ is defined as,
\begin{equation}
L = \max\left(0, \mathcal{D}(f^u_1,f^u_2) - \mathcal{D}(f^u_1,f_v) + \alpha\right)
\label{Eq:tl-enrollment}
\end{equation}
,where $\mathcal{D}$ is Euclidean distance and $\alpha$ is a hyper-parameter. 


\subsection{Feature Extraction}

\begin{figure}
\begin{center}
   \includegraphics[width=1\linewidth]{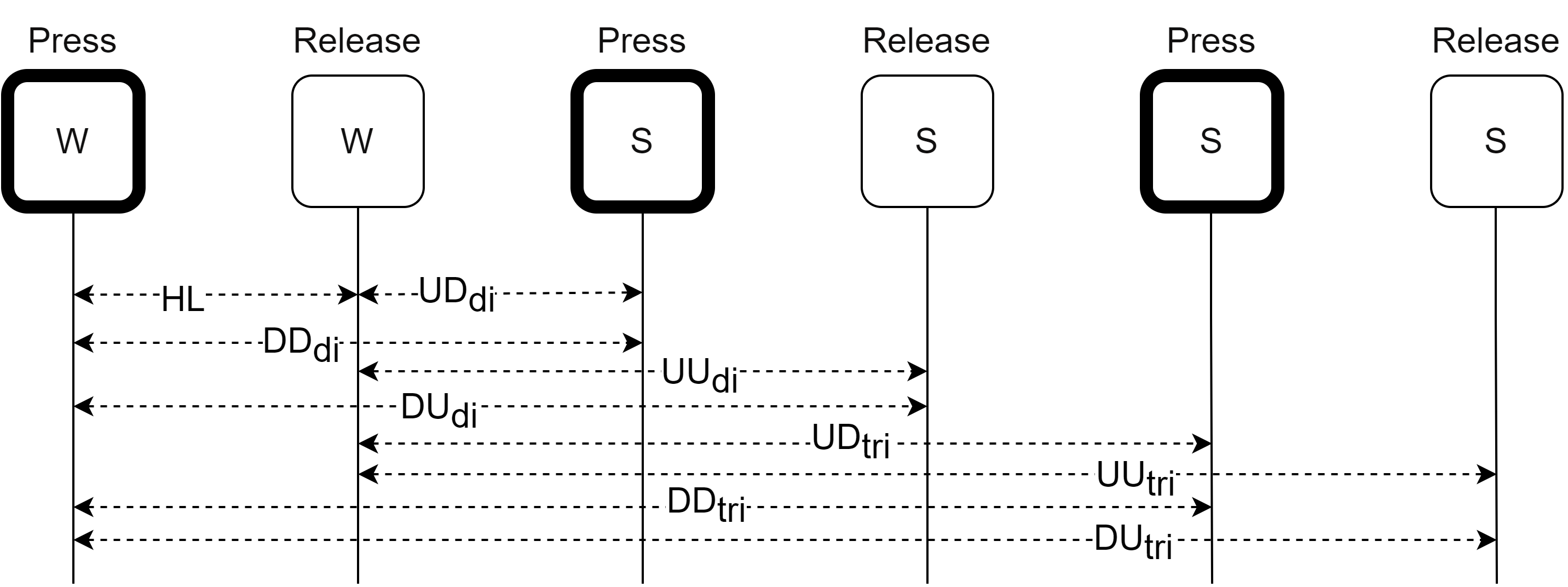}
\end{center}
   \caption{Keystroke Feature Extraction: Di-Gram ($DU_{di}$, $UD_{di}$, $DD_{di}$, $UU_{di}$), Tri-Gram ($DU_{tri}$, $UD_{tri}$, $DD_{tri}$, $UU_{tri}$), and Hold Latency (HL) Illustration.}
\label{fig:long}
\label{fig:Example of the keystroke features extracted from databases.}
\end{figure}

\paragraph{Keystroke features:} We extract Di-Gram and Tri-Gram features (See Fig.~\ref{fig:Example of the keystroke features extracted from databases.}) as keystroke features. Di-Grams consider two consecutive key presses and Tri-Grams consider three consecutive key presses to extract features. We extract time between different events as features, for example UD represents the time between a key up event (U) and key down event (D). Similarly we extract DD, DU and UU times. Hold Latency (HL) and ASCII code together gives us the 10 keystroke features.

$[ HL, DU_{di}, UD_{di}, DD_{di}, UU_{di}, DU_{tri}, UD_{tri},$
$ DD_{tri}, UU_{tri}, ASCII]$ 

ASCII code is normalized to the range [0,1] and time-based features are represented in seconds. HuMIdb dataset doesn't differentiate between press and release time, therefore only has a subset of features; $[DD_{di}, DD_{tri}, ASCII]$.

\paragraph{IMU features:} To incorporate IMU data, the first and second-order derivatives are computed from the raw $x$, $y$, $z$ values, as well as the Fast Fourier Transform (FFT) \cite{deb2019actions}. For the FFT values, only the absolute values are taken after FFT calculation. Ultimately, for each sensor data type, a 12-dimensional vector is generated for each timestep; 

$\left [ x, y, z, {x}', {y}', {z}', {x}'', {y}'', {z}'', fft\left ( x \right ), fft\left ( y \right ), fft\left ( z \right ) \right ]$

Therefore after combining all IMU sensor data type features, a 36-dimensional vector is obtained. All values are normalised in the range of [0,10].



\begin{figure}
\begin{center}
   \includegraphics[width=0.65\linewidth]{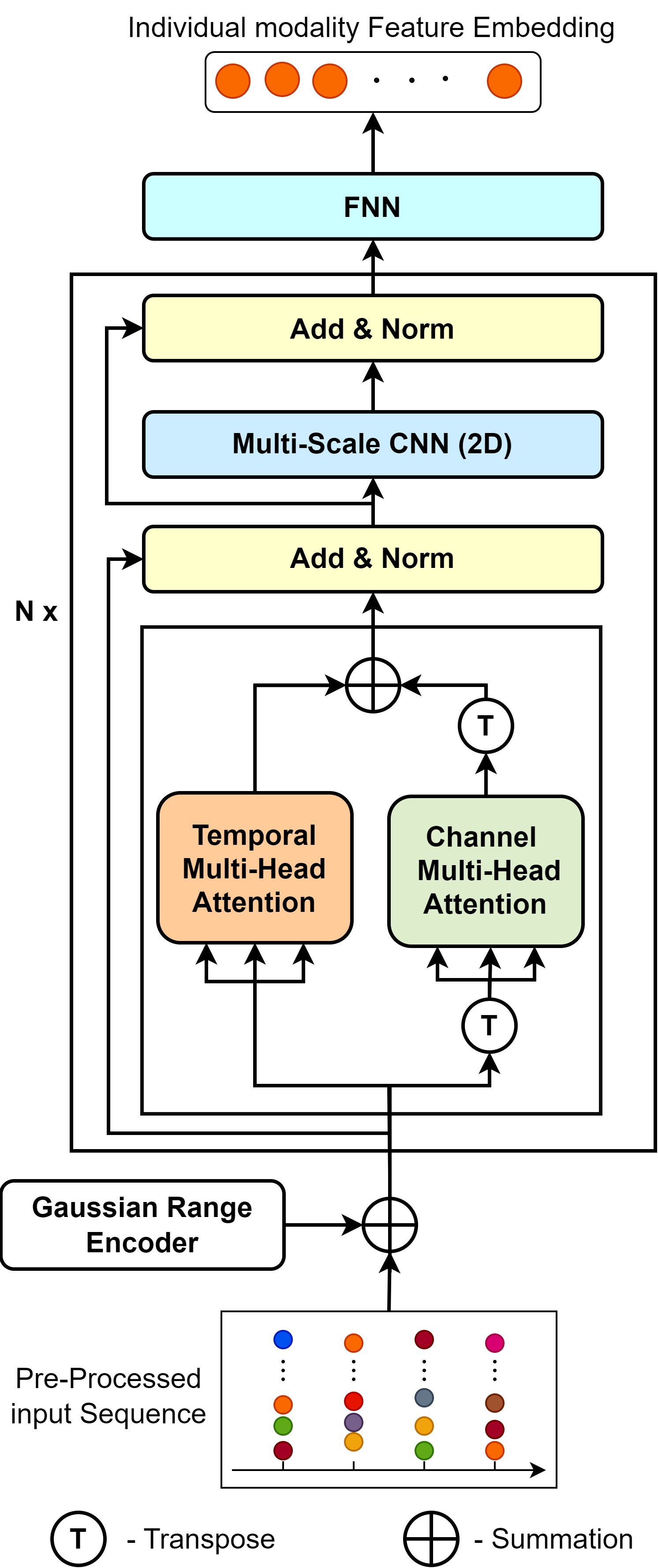}
\end{center}
   \caption{The high level architecture of Spatio-Temporal Dual Attention Transformer (STDAT) consisting of Gaussian Range Encoder as the Positional Encoder, dual attention with temporal and channel Multi-Head Attentions, Multi-Scale 2D CNN and FNN which generates the individual modality feature embedding.}
\label{fig:Transformer component}
\label{fig:onecol}
\end{figure}

\subsection{Spatio-Temporal Dual Attention Transformer}

The success of behavioural biometric authentication models relies on identifying the unique behavioural patterns of individual users. We can use keystroke dynamics over time to extract unique patterns to achieve this. 

Our new Spatio-Temporal Dual Attention Transformer (STDAT) improves upon the Vanilla Transformer's encoder part \cite{vaswani2017attention} to extract user-specific behavioural patterns. However, unlike previous models, \cite{stragapede_mobile_2022} and ~\cite{stragapede2022typeformer}, which utilized two transformers as in~\cite{Li_Cui_Wang_Zhang_Chen_Wu_2021} to extract time-over-channel and channel-over-time features, we use a single transformer with dual attention: attention over the temporal axis and attention over channel axis. This approach enables STDAT to focus on the relevant keystroke features over time, extracting unique behavioural patterns for individual users. For an overview of STDAT, refer to Fig.~\ref{fig:Transformer component}.


Suppose the pre-processed input to the STDAT is $X \in R^{N \times M}$ where $N$ is the sequence length and $M$ is the number of features. Since the positional details samples are imperative in generating discriminative feature for each individual user, we first add the positional encoding to the input $X$ using the Gaussian Range Encoder (GRE) proposed in \cite{Li_Cui_Wang_Zhang_Chen_Wu_2021}. GRE is a learnable range-based positional encoding that encodes ranges of positions instead of just one point in contrast to other positional encodings, like absolute and relative positional encoding. With GRE, a single sequence position is defined as a normalized Probability Density Function vector of K uni-variate Gaussian Distributions. This vector is then multiplied by a K learnable range embedding to create the final GRE. Let $G \in R^{N \times M}$ be the Gaussian positional encoding, then the range-biased input is defined as:
\begin{equation}
\Bar{X} = X + G
\label{Eq:1}
\end{equation}

$\Bar{X}$ is then fed into the new dual attention block comprising two Multi-Head Attention (MHA) modules: Temporal-MHA and Channel-MHA. Temporal-MHA analyzes input data over time to extract information from the original sequence, while Channel-MHA examines input data across different channels using the transposed input. Each attention module processes its input sequence and calculates a set of attention weights for each element in the sequence (time-wise or channel-wise), showing the element's importance toward the final output. These attention weights are then used to calculate a weighted sum of the input sequence, which forms the output of the multi-head attention module. If the output of Temporal-MHA is $V_{T-MHA} \in R^{N \times M}$ and the output of Channel-MHA is $V_{C-MHA} \in R^{M \times N}$, then the final output of the dual attention block is $V = V_{T-MHA} + (V_{C-MHA})^T$. $V$ contains discriminative features derived considering both temporal and channel patterns. 
 
\begin{table*}[tbp]
\centering
\begin{tabular}{|l|l|l|l|l|} 
\hline
\multicolumn{1}{|c|}{\textbf{Dataset}} & \multicolumn{1}{c|}{\textbf{Subjects}} & \multicolumn{1}{c|}{\textbf{Sessions}} & \multicolumn{1}{c|}{\textbf{Actions}} & \multicolumn{1}{c|}{\textbf{Modalities}} \\ 
\hline
Aalto DB~\cite{10.1145/3338286.3340120} &  $\sim$260,000 & 15 & Typing & Keystroke \\ 
\hline
HMOG DB~\cite{7349202} & 100 & 24 & \begin{tabular}[c]{@{}l@{}}Reading, Typing, Map navigation\end{tabular} & \begin{tabular}[c]{@{}l@{}}Keystroke, Raw touch event, IMU \end{tabular} \\ 
\hline
HuMIdb~\cite{ACIEN2021104058} & 599 & 1-5 & \begin{tabular}[c]{@{}l@{}}Typing, Swipe, Tap, hand gesture,\\finger writing\end{tabular} & \begin{tabular}[c]{@{}l@{}}Keystroke, Touch, IMU, Light,  GPS, \\ WiFi, Bluetooth, Orientation,\\Proximity, Microphone\end{tabular} \\
\hline
\end{tabular}
\caption{An overview outlining the number of subjects and sessions, as well as the various types of actions and modalities available within the three datasets (Aalto DB, HMOG DB, and HuMIdb) used in the study.}
\label{tab:Datasets}
\end{table*}

The dual attention block is followed by a residual connection \cite{residualconnection} (Add) and a layer normalization~\cite{ba2016layer} (LayerNorm) layer. This layer adds $V$ and $\Bar{X}$ and normalizes values to produce $\Bar{V}$. $\Bar{V}$ is then fed into a Multi-scale 2D Convolutional Neural Network (M2D-CNN) block consisting of several 2D convolution layers with different kernel sizes, each followed by Batch Normalisation layer~\cite{ioffe2015batch}, Dropout Layer~\cite{hinton2012improving} and ReLU activation function. 2D convolutional enables extracting features from $\Bar{V}$ considering both temporal and time axes, ultimately increasing the discriminativeness of the output feature embedding. The output of the M2D-CNN block is then passed to another Add and Norm layer, which adds it to $\Bar{V}$ and normalizes it to produce the final output of the dual attention block. 

The STDAT model is made up of multiple dual attention blocks, with each block taking the output of the previous one and generating a new feature embedding for the next block. The last block's output is then passed through a Fully Connected Neural Network (FNN) consisting of two fully connected layers using ReLU activation functions. The FNN's output is the feature embedding for the Keystroke dynamic input.

\section{Experimental Study}

\begin{table*} [tbp]
\centering
\begin{tabular}{|c|l|c|c|c|c|c|} 
\hline
\textbf{Modality} & \multicolumn{1}{c|}{\textbf{Dataset}} & \textbf{EER} & \textbf{Usability} & \textbf{TCR} & \textbf{FRWI} & \textbf{FAWI} \\ 
\hline
\multirow{3}{*}{\rotcell{Keystroke}} & Aalto DB & 1.80 & 0.99 & 12.70 & 0.01 & 0.51 \\ 
\cline{2-7}
 & HMOG DB & 5.10 & 0.95 & 216.87 & 0.23 & 8.39 \\ 
\cline{2-7}
 & HuMIdb & 12.04 & 0.91 & 24.92 & 0.03 & 1.16 \\ 
\hline
\multirow{2}{*}{\rotcell{\begin{tabular}[c]{@{}c@{}}Keystroke+IMU\end{tabular}}} & HMOG DB & 3.62 & 0.97 & 161.72 & 0.66 & 6.12 \\ 
\cline{2-7}
 & HuMIdb & 2.95 & 0.99 & 17.19 & 0.00 & 0.60 \\
\hline
\end{tabular}

\caption{Performance of BehaveFormer Across Datasets. The top three rows correspond to BehaveFormer trained solely on keystroke data (K), while the last two rows show its performance when trained on both keystroke and IMU data (K+IMU). The reported metrics include EER and Usability (as fractions), TCR (in seconds), and FRWI and FAWI (in minutes).}
\label{tab:Overall results}
\end{table*}

We use three publicly available datasets which are Aalto DB, HMOG DB and HuMIdb. All three datasets contain keystroke data but only HMOG DB and HuMIdb contain touchstroke and IMU data. An overview of these datasets is provided in Table~\ref{tab:Datasets}.


\begin{figure*} [h]
\centering
\begin{subfigure}{.4\textwidth}
  \centering
  \includegraphics[width=1\linewidth]{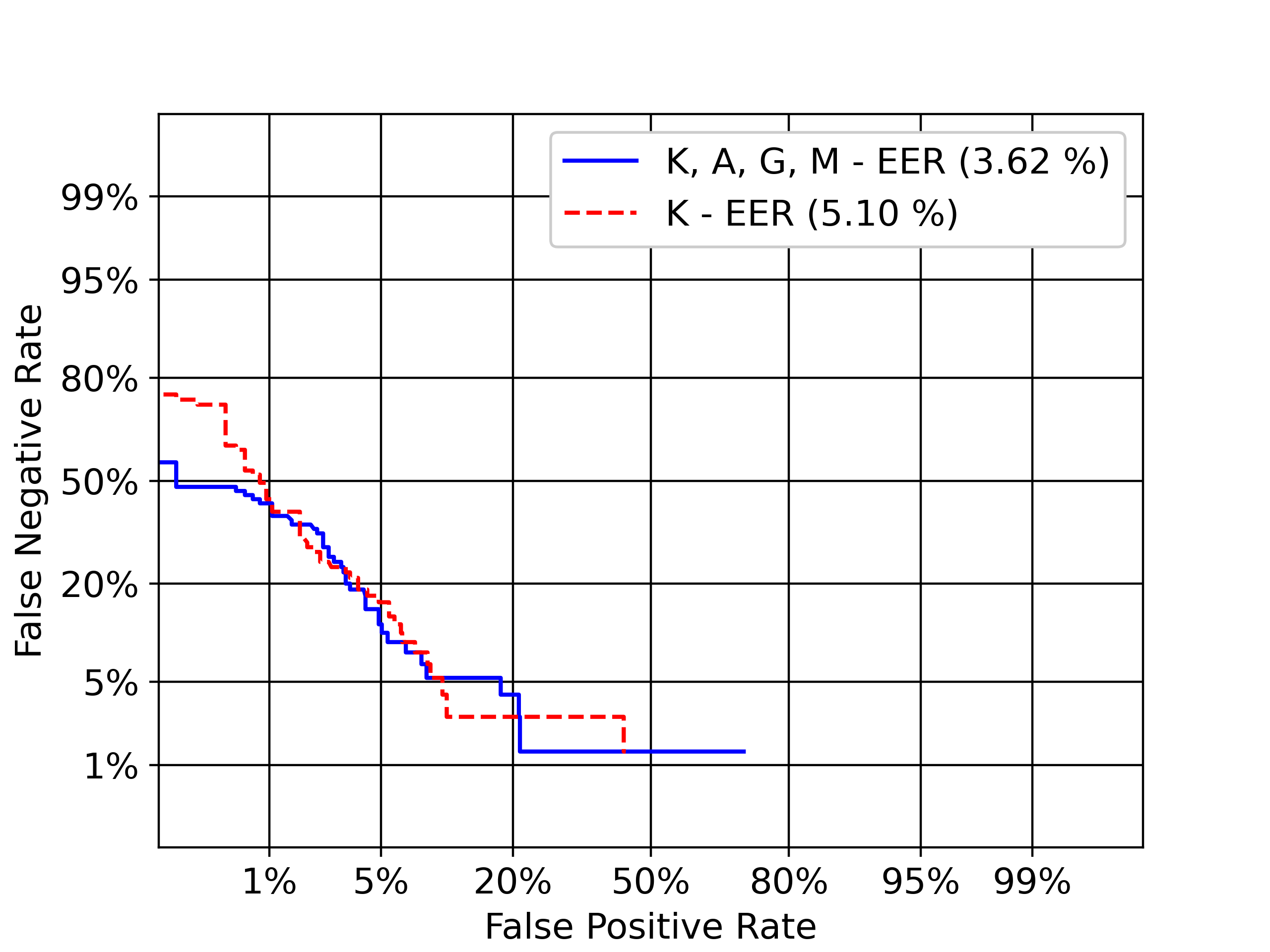}  
  \caption{HMOG DB}
  \label{fig:sub-first}
\end{subfigure}
\begin{subfigure}{.4\textwidth}
  \centering
  \includegraphics[width=1\linewidth]{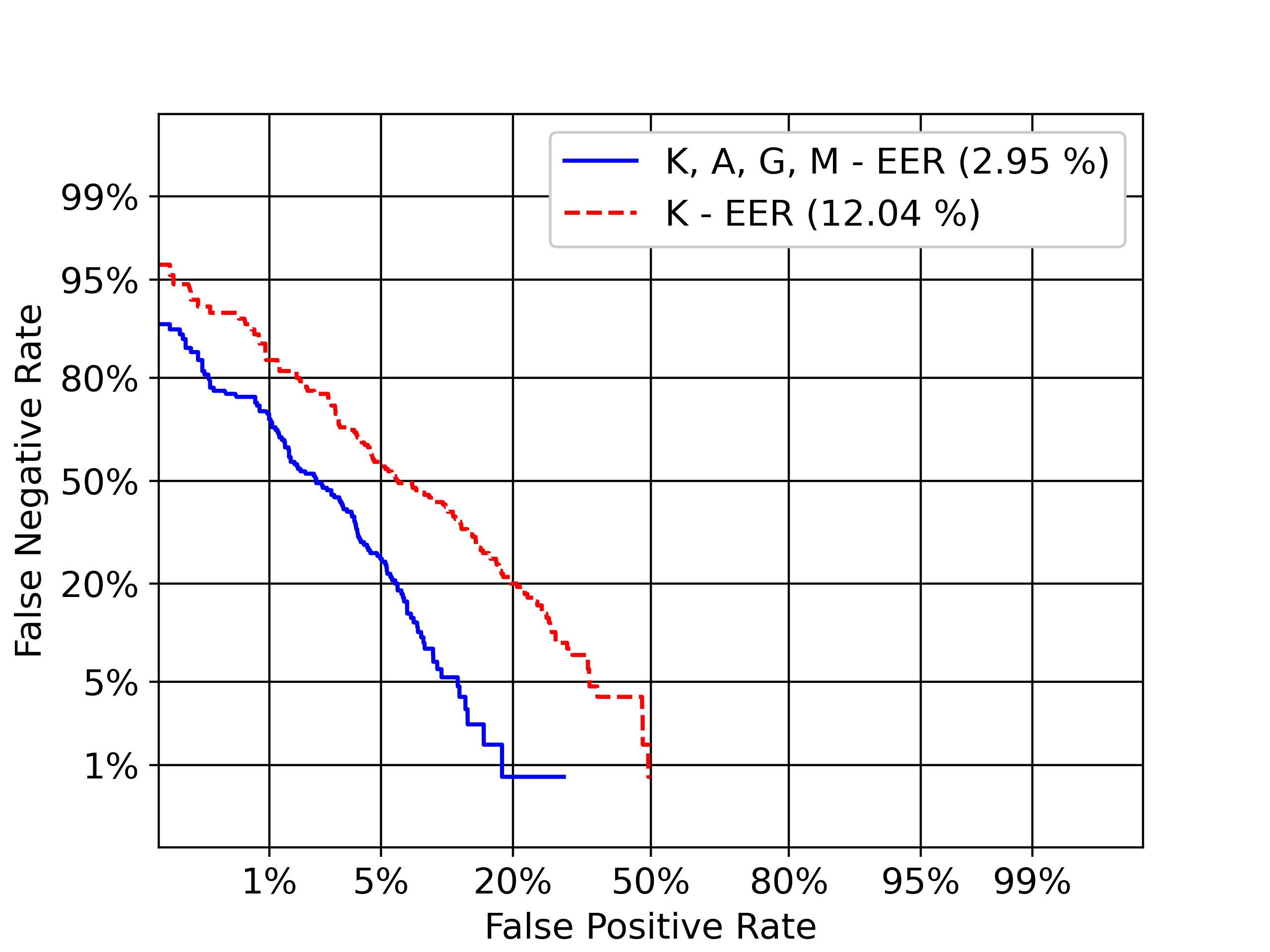}  
  \caption{HuMIdb}
  \label{fig:sub-fourth}
\end{subfigure}
\caption{The DET curves generated for the keystroke only scenario and keystroke combined with IMU data (best performing IMU data combination) scenario we consider for BehaveFormer.}
\label{fig:det_curves.}
\end{figure*}

\begin{figure*} [h]
  \subfloat[Keystroke only - Aalto DB]{
	\begin{minipage}[c][1\width]{
	   0.3\textwidth}
	   \centering
	   \includegraphics[width=1\textwidth]{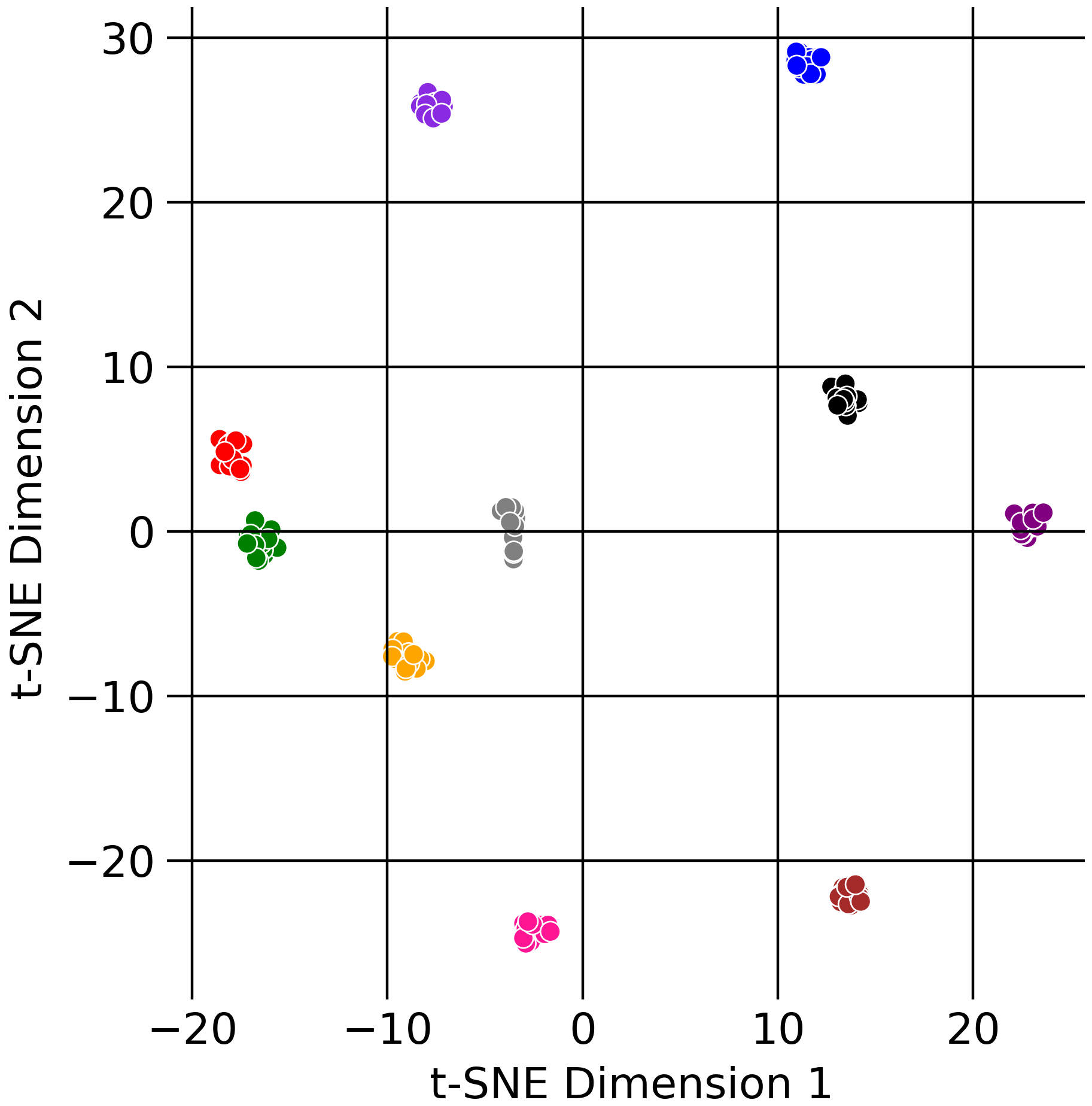}
	\end{minipage}}
 \hfill 	
  \subfloat[Keystroke with IMU data - HMOG DB]{
	\begin{minipage}[c][1\width]{
	   0.3\textwidth}
	   \centering
	   \includegraphics[width=1\textwidth]{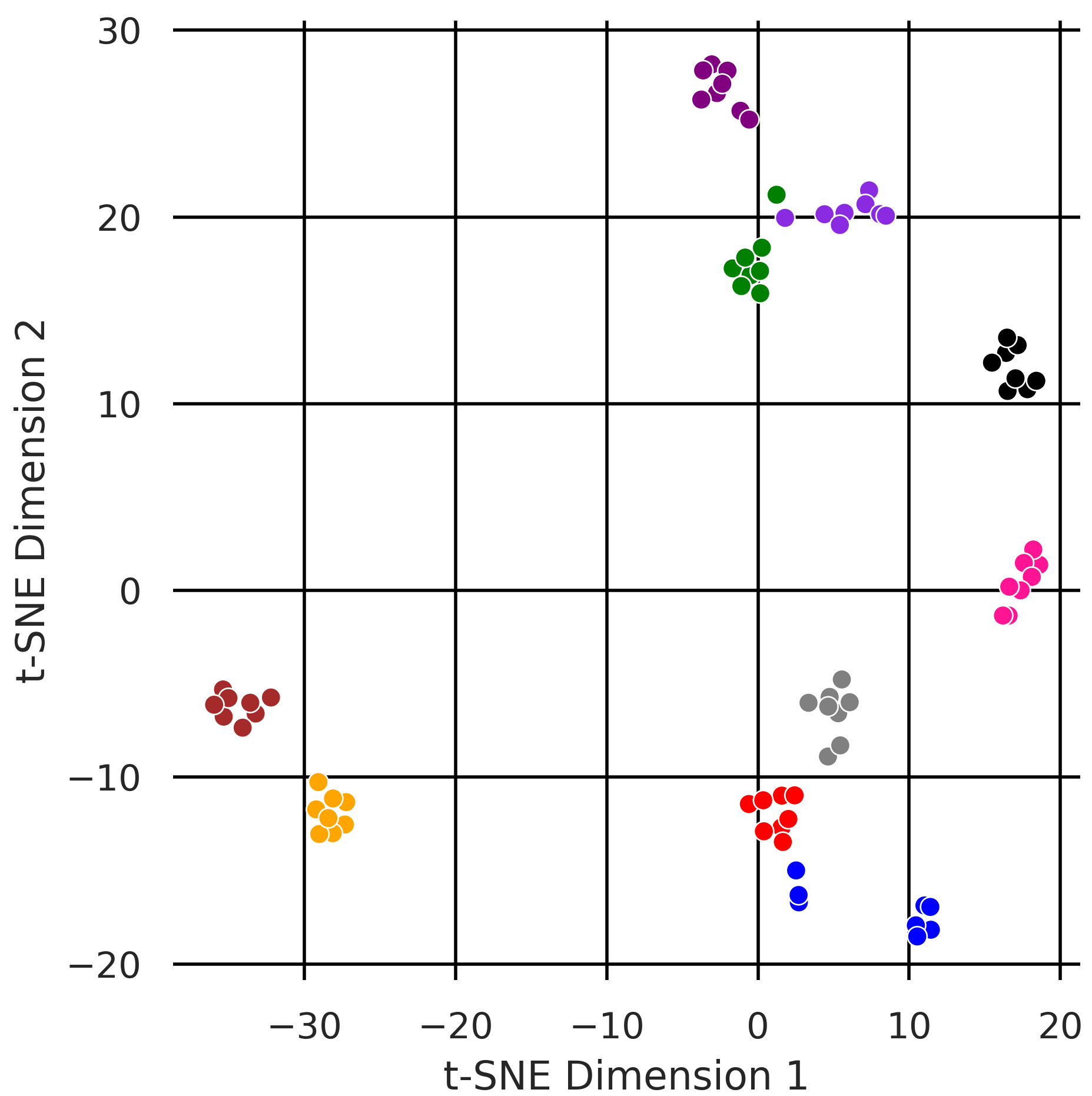}
	\end{minipage}}
 \hfill	
  \subfloat[Keystroke with IMU data - HuMIdb]{
	\begin{minipage}[c][1\width]{
	   0.3\textwidth}
	   \centering
	   \includegraphics[width=1\textwidth]{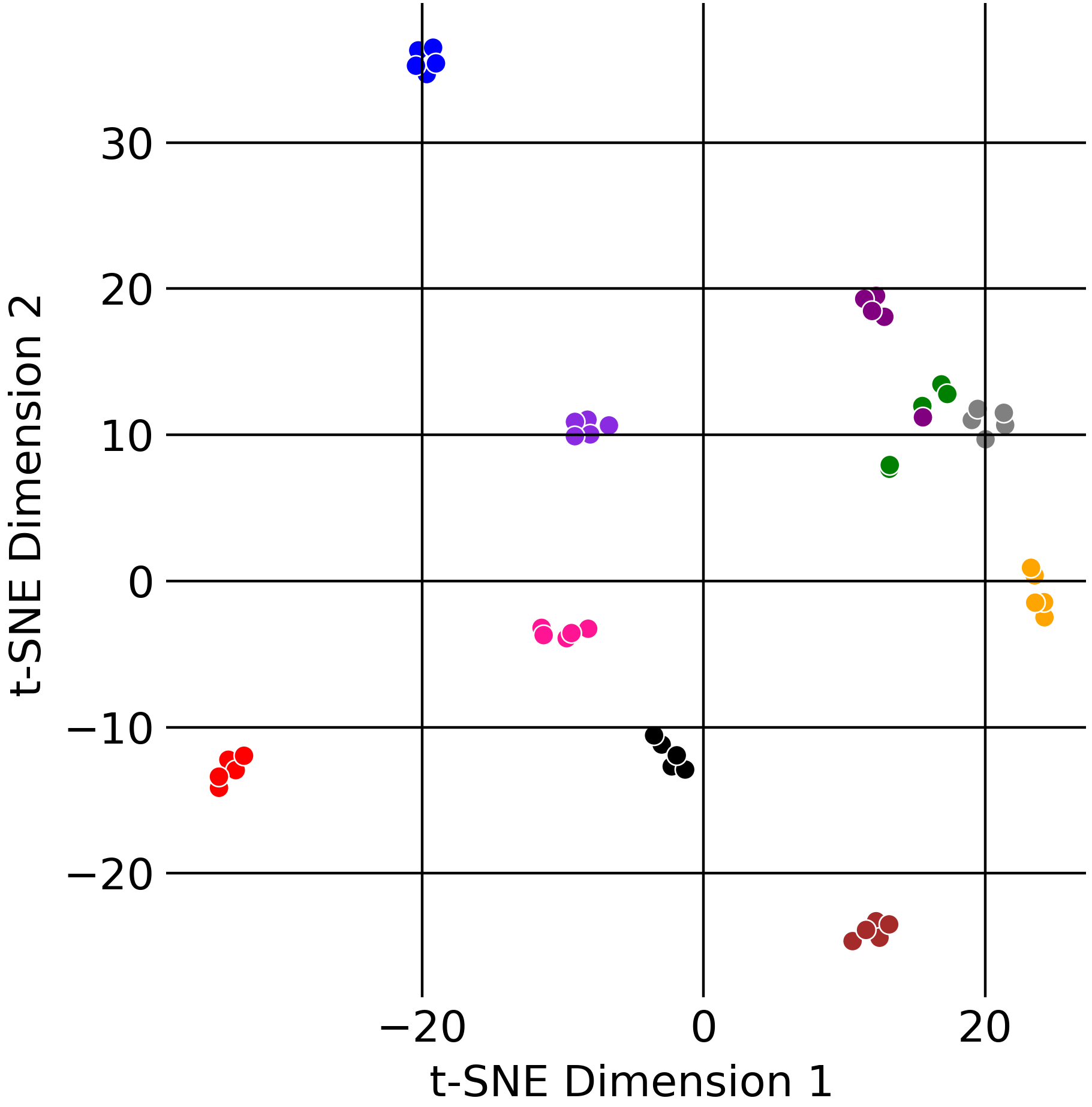}
	\end{minipage}}
\caption{2D graphical visualisation of the latent space through tSNE considering 10 subjects (different colours represent different subjects) \cite{hinton2008visualizing}. Used parameters: $perplexity = 14 (Aalto DB), 7 (HMOG DB), 4 (HuMIdb), init = 'pca', n\_iter = 1000, n\_components=2, learning\_rate='auto'.$}
\label{fig:tsne_graphs}
\end{figure*}

\subsection{Pre-processing}
 



In data pre-processing, we filter out users based on data availability and then split the remaining data into training, testing, and validation sets.

Aalto DB had 60,000 users after filtering for data availability. These users were randomly divided into three sets: training (30,000), testing (1,000), and validation (400) sets. HMOG DB had only 99 users left after filtering for IMU data availability. These users were randomly divided into training (69), testing (15), and validation (15) sets. For HuMIdb, 428 users were selected after removing those without keystrokes. These users were randomly divided into training (328), testing (65), and validation (35) sets.

The IMU data is synchronized with keystroke sequences during the final pre-processing step. This is done by averaging the IMU data within a dynamic time period for each keystroke sequence. The resulting IMU data sequence is then standardized to a fixed size of 100.




\subsection{Implementation Details}
The BehaveFormer is implemented using PyTorch \cite{whitepaperpytorch} using Adam optimizer \cite{adamopt} and a learning rate value of 0.001. In addition, the $\alpha$ hyper-parameter in the training objective of BehaveFormer is set to 1.

Next, we will provide details about the optimal hyperparameters of the proposed model. To begin with, we utilized 20 Gaussian distributions for the GRE. For keystroke-STDAT, we used 6,5 and 5 dual attention blocks for Aalto DB, HMOG DB, and HuMIdb, respectively. On the other hand, IMU-STDAT consisted of five dual attention blocks for both HMOG and HuMidb datasets. The Multi-scale 2D CNN used in both types of STDATs had three convolutional layers with kernel sizes of 1, 3, and 5. The final feature embedding size of both STDATs was 64. For the Temporal-MHA and Channel-MHA, Keystroke-STDAT used 5 and 10 attention heads for both Aalto and HMOG DB datasets, while three heads were used in Temporal-MHA and ten heads in Channel-MHA in HuMIdb. In contrast, IMU-STDAT utilized 6 and 10 heads for the Temporal-MHA and Channel-MHA in both HMOG DB and HuMIdb implementations.

\subsection{Verification Protocol}
The evaluation process employs the enrollment-verification approach \cite{Acien2022}. The models are trained using a distinct set of users and assessed using the remaining users. A proportion of each test user's data is allocated for enrollment, while the remaining data is used for verification. During the evaluation process of a user, all other testing users are considered imposters, while each metric is computed for every user in the test set, with an average taken across all users. 

The process of subject authentication involves comparing the enrollment feature embedding $f_{u,e}$ belonging to a particular subject $u$ in the test set with a verification feature embedding $f_{v,a}$. The comparison is made between either the same subject (genuine match $v = u$) or another subject (impostor $v\neq u$). To compute the test score, we calculate the average Euclidean distances between each $f_{u,e}$ and $f_{v,a}$ as shown in Equation~\ref{Eq:verification-enrollment}.
\begin{equation}
s_{u,v}^{a}= \frac{1}{E}\sum_{e=1}^{E}\left \|  f_{u,e} - f_{v,a} \right \|
\label{Eq:verification-enrollment}
\end{equation}
where $E$ is the number of enrollment samples and $a$ is the verification sample of subject $v$.



\subsection{Performance of BehaveFormer Across Datasets}

First, we conducted a performance evaluation of BehaveFormer on multiple datasets using various metrics. In this experiment, we trained two versions of BehaveFormer: one using only Keystroke data and the other using both Keystroke and IMU data. Overall results of BehaveFormer's performance on Aalto DB, HMOG DB, and HuMIdb datasets are displayed in Table~\ref{tab:Overall results}. BehaveFormer has demonstrated high usability and low EER across all datasets. Moreover, the version of BehaveFormer trained with both keystroke and IMU data outperformed the version trained solely on keystroke data, indicating that combining keystroke data with IMU data has improved the discriminativeness of the learned feature embedding. 


\begin{table*} [tbp]
\centering
\begin{tabular}{|c|l|l|c|c|c|} 
\hline
\multirow{2}{*}{\textbf{Modality}} & \multicolumn{1}{c|}{\multirow{2}{*}{\textbf{Study}}} & \multicolumn{1}{c|}{\multirow{2}{*}{\textbf{Model Type}}} & \multicolumn{3}{c|}{\textbf{EER (\%)}} \\ 
\cline{4-6}
 & \multicolumn{1}{c|}{} & \multicolumn{1}{c|}{} & \multicolumn{1}{l|}{\textbf{AaltoDB}} & \textbf{HMOG DB} & \multicolumn{1}{l|}{\textbf{HuMIdb}} \\ 
\hline
\multirow{5}{*}{\rotcell{Keystroke}} & TypeNet & LSTM & 8.00* & 8.67 & 12.40 \\ 
\cline{2-6}
 & HuMINet & LSTM & 15.10 & 13.37 & 12.19* \\ 
\cline{2-6}
 & DuoNet & LSTM & 12.51 & 36.21 & 12.19* \\ 
\cline{2-6}
 & TypeFormer & Transformer & 3.15* & 17.48 & 20.76 \\ 
\cline{2-6}
 & \textbf{BehaveFormer} & \textbf{Transformer} & \textbf{1.80} & \textbf{5.10} & \textbf{12.04} \\ 
\hline
\multirow{3}{*}{\rotcell{Keystroke+IMU}} & HuMINet & LSTM & – & 19.97 & 3.96* \\ 
\cline{2-6}
 & DuoNet & LSTM & – & 46.47 & 7.58* \\ 
\cline{2-6}
 & \textbf{BehaveFormer} & \textbf{Transformer} & \textbf{ –} & \textbf{3.62} & \textbf{2.95} \\
\hline
\end{tabular}
\caption{Comparison of BehaveFormer with state-of-the-art keystroke dynamics models. The top six rows correspond to BehaveFormer compared with models solely trained on keystroke data. The bottom three rows correspond to BehaveFormer compared to models trained on keystroke and IMU data. The results taken from the respective original papers are marked by an asterisk (*), while others are from our implementations of the existing models.}
\label{tab:comparative_study}
\end{table*}


\begin{table*} [tbp]
\centering
\small
\resizebox{\linewidth}{!}{%
\begin{tabular}{|c|l|c|c|c|c|c|c|c|c|c|c|c|c|} 
\hline
\multicolumn{2}{|c|}{\multirow{2}{*}{\begin{tabular}[c]{@{}c@{}}\textbf{}\\\textbf{Model}\end{tabular}}} & \multicolumn{4}{c|}{\textbf{Aalto DB}} & \multicolumn{4}{c|}{\textbf{HMOG DB}} & \multicolumn{4}{c|}{\textbf{HuMIdb}} \\ 
\cline{3-14}
\multicolumn{2}{|c|}{} & \textbf{Usab.} & \textbf{TCR} & \textbf{FRWI} & \textbf{FAWI} & \textbf{Usab.} & \textbf{TCR} & \textbf{FRWI} & \textbf{FAWI} & \textbf{Usab.} & \textbf{TCR} & \textbf{FRWI} & \textbf{FAWI} \\ 
\hline
\multirow{5}{*}{\rotcell{K}} & TypeNet & 0.92 & 16.94 & 0.09 & 1.16 & 0.93 & 321.68 & 0.49 & 11.05 & 0.86 & 23.55 & 0.01 & 1.04 \\ 
\cline{2-14}
 & TypeFormer & 0.89 & 18.31 & 0.11 & 1.40 & 0.83 & 377.83 & 1.54 & 14.99 & 0.82 & 31.32 & 0.05 & 1.43 \\ 
\cline{2-14}
 & HuMINet & 0.89 & 50.47 & 0.11 & 71.26 & 0.87 & 267.49 & 0.73 & 11.59 & 0.89 & 25.16 & 0.03 & 1.13 \\ 
\cline{2-14}
 & DuoNet & 0.91 & 47.27 & 0.09 & 52.95 & 0.84 & 531.92 & 0.53 & 18.49 & 0.90 & 25.92 & 0.02 & 0.97 \\ 
\cline{2-14}
 & BehaveFormer &\textbf{ 0.99} &\textbf{ 12.70} & \textbf{0.01} & \textbf{0.51} & 0.95 & 216.87 & \textbf{0.23} & 8.39 & 0.91 & 24.92 & 0.03 & 1.16 \\ 
\hline
\multirow{3}{*}{\rotcell{\begin{tabular}[c]{@{}c@{}}K+IMU\end{tabular}}} & HuMINet & \_ & \_ & \_ & \_ & 0.82 & 394.14 & 0.93 & 15.57 & 0.92 & 20.31 & 0.02 & 0.96 \\ 
\cline{2-14}
 & DuoNet & \_ & \_ & \_ & \_ & 0.61 & 615.52 & 1.00 & 18.61 & 0.96 & \textbf{16.57} & 0.02 & 0.70 \\ 
\cline{2-14}
 & BehaveFormer & \_ & \_ & \_ & \_ &\textbf{ 0.97} &\textbf{ 161.72 }& 0.66 & \textbf{6.12} & \textbf{0.99} & 17.19 & \textbf{0.00} &\textbf{ 0.60} \\
\hline
\end{tabular}
}

\caption{Evaluation of keystroke dynamics with CA evaluation metrics. The reported metrics are Usability (as a fraction), TCR (in seconds), and FRWI and FAWI (in minutes). Best result for each dataset for each evaluation criteria is highlighted in \textbf{bold}.}
\label{tab:CA-comparative-study}
\end{table*}

\begin{table*}[tbp]
\centering
\begin{tabular}{|l|c|c|c|c|c|c|c|c|c|c|} 
\hline
\multirow{2}{*}{\textbf{Modalities}} & \multicolumn{5}{c|}{\textbf{HMOG DB }} & \multicolumn{5}{c|}{\textbf{HuMIdb }} \\ 
\cline{2-11}
 & \textbf{EER (\%)} & \textbf{Usability} & \textbf{TCR} & \textbf{FRWI} & \textbf{FAWI} & \textbf{EER (\%)} & \textbf{Usability} & \textbf{TCR} & \textbf{FRWI} & \textbf{FAWI} \\ 
\hline
K & 5.10 & 0.95 & 216.87 & \textbf{0.23} & 8.39 & 12.04 & 0.91 & 24.92 & 0.03 & 1.16 \\ 
\hline
K, A & 4.95 & 0.95 & 234.23 & 0.90 & 8.47 & 6.26 & 0.96 & 20.77 & 0.02 & 0.77 \\ 
\hline
K, G & 4.52 & 0.96 & 188.60 & 0.59 & 6.61 & 6.78 & 0.96 & 21.25 & 0.02 & 0.90 \\ 
\hline
K, M & 4.86 & 0.96 & 204.50 & 0.48 & 6.92 & 4.39 & 0.97 & 17.87 & \textbf{0.00} & 0.64 \\ 
\hline
K, A, G & 3.81 & \textbf{0.97} & 166.74 & 0.66 & 6.33 & 4.38 & 0.97 & 19.10 & \textbf{0.00} & 0.78 \\ 
\hline
K, A, M & 4.0 & \textbf{0.97} & 236.87 & 0.66 & 7.76 & 4.81 & 0.96 & 20.12 & \textbf{0.00} & 0.75 \\ 
\hline
K, G, M & 4.19 & 0.96 & \textbf{159.51} & 0.31 & \textbf{5.67} & 7.91 & 0.93 & 25.83 & \textbf{0.00} & 0.97 \\ 
\hline
K, A, G, M & \textbf{3.62} & \textbf{0.97} & 161.72 & 0.66 & 6.12 & \textbf{2.95} & \textbf{0.99} & \textbf{17.19} & \textbf{0.00} & \textbf{0.60} \\
\hline
\end{tabular}


\caption{Ablation study for varying data modalities used for training. The modalities tested include Keystroke (K), Accelerometer (A), Magnetometer (M) and Gyroscope (G).The reported metrics are EER and Usability (as fractions), TCR (in seconds), and FRWI and FAWI (in minutes). Best result for each column is highlighted in \textbf{bold}.}
\label{tab:PROPOSED MODEL RESULTS FOR HMOG}
\end{table*}

To further assess the impact of adding IMU data on BehaveFormer's performance, we compared the best-performing models with and without IMU data. We plotted their Detection Error Trade-off (DET) curves in Fig.~\ref{fig:det_curves.}. The curves show the global FAR, FRR, and EER values, with the local EER values listed in the legend. By comparing the two curves, it's clear that adding IMU data improved BehaveFormer's performance.

Next, we demonstrate that BehaveFormer can learn unique feature embedding for different subjects. Using t-SNE graphs, we visualized the feature embedding space of BehaveFormer for ten subjects across three datasets, as shown in Fig.\ref{fig:tsne_graphs}. In all datasets, the embeddings for each subject were distinct and well-separated. For instance, the mean Silhouette score for AaltoDB, HMOG DB, and HuMIdb was 0.91, 0.69, and 0.75, respectively. High Silhouette scores indicate low intra-class and high inter-class variability, meaning BehaveFormer can learn unique feature embeddings for each subject.

\subsection{Comparative Study}

We compare the performance of proposed BehaveFormer with the following state-of-the-art behavioural biometric-based CA systems.
\begin{itemize}
\itemsep0em   \item\textbf{TypeNet}~\cite{Acien2022} - TypeNet is a free-text keystroke biometric CA system using a deep recurrent neural network and trained with triplet loss for authentication and identification at a large scale.
  \item \textbf{HuMINet}~\cite{Stragapede2022} - A multimodal architecture with two LSTM layers in each model. These models are combined at the score level using either a weighted or non-weighted method. 
  \item \textbf{DuoNet}~\cite{stragapede2022mobile} - This system uses the same multimodal structure as HuMINet but also goes through an extra training stage. During this second phase, the feature embeddings from the initial training process are utilized to train a completely new model. 
  \item \textbf{TypeFormer}~\cite{stragapede_mobile_2022} - Follows a modified transformer architecture with two encoders: Temporal and Channel Modules. Each encoder consists of a Multi-Head Self-Attention mechanism and a Multi-Scale Keystroke CNN to extract temporal and channel features. 
\end{itemize}
This experiment compares BehaveFormer and other state-of-the-art behavioural biometric-based CA models. We tested two scenarios: training solely with keystroke data and training with both keystroke and IMU data. The results are shown in Table~\ref{tab:comparative_study}. When trained with only keystroke data, BehaveFormer outperforms the existing for both Aalto DB and HuMIdb by achieving an EER of 1.80\% and 12.04\%, respectively, which is an improvement of 29.3\% and 1.23\%. Similarly, BehaveFormer, trained with the keystroke and IMU data, outperforms the existing models trained with both keystroke dynamics. For example, BehaveFormer achieves the best EER of 2.95\% on HuMIdb, which is an improvement of 25.76\% over the state-of-the-art.

To better understand how these models perform in a CA environment, we evaluated them using specific CA evaluation metrics, such as Usability, TCR, FRWI, and FAWI \cite{10.1109/TPAMI.2007.1010,10.1109/TC.2016.2622262}. Table~\ref{tab:CA-comparative-study} presents the results of this evaluation. We observe that BehaveFormer maintains its superior performance across all CA evaluation metrics.

\subsection{Transfer Learning}
The small size of available datasets often limits behavioural biometrics research. One way to overcome this limitation is through transfer learning which involves training a model on a large dataset for a specific task and then transferring the learned features to a second model trained on a different task. To demonstrate transfer learning in behavioural biometrics, we compared the performance of a BehaveFormer model trained from scratch with the HMOG dataset versus a BehaveFormer model pre-trained on Aalto DB, which was then fine-tuned on HMOG DB. The model trained from scratch achieved an EER of 5.10\%, a Usability score of 0.95, a TCR of 216.87 seconds, an FRWI of 0.23 min, and a FAWI of 8.39 min. On the other hand, the BehaveFormer model using transfer learning achieved an EER of 4.48\%, a Usability score of 0.95, a TCR of 220.88 seconds, an FRWI of 0.55 min, and a FAWI of 7.46 min. The use of transfer learning resulted in a 12.16\% improvement in EER, highlighting its potential in overcoming small dataset issues in behavioural biometrics research.

\subsection{Ablation Study}

Finally, we conducted an ablation study to reinforce our selection of background sensor data from the IMU stream. During this process, we tested various combinations of data modalities such as Keystroke (K), Accelerometer (A), Magnetometer (M), and Gyroscope (G). We analyzed HMOG and HuMIdb datasets and summarized the results in Table \ref{tab:PROPOSED MODEL RESULTS FOR HMOG}. Our research revealed that combining keystroke data with all three types of IMU data (K, A, G, M) produced the best results, with an EER of 3.62\% and 2.95\% for the two datasets, respectively. Furthermore, our findings indicate that using all IMU sensors consistently leads to high performance for the CA evaluation metrics. 

\section{Discussion}

The proposed BehaveFormer model outperforms state-of-the-art CA systems in both keystroke-only and IMU-enhanced scenarios. The learned feature space achieves high separation among different identities. The ablation study further demonstrates the positive impact of using IMU data (especially all three types of IMU data we have used in the research) on the overall system. Therefore, this shows how more discriminative features can be extracted with the use of IMU data for keystroke dynamics. While BehaveFormer outperforms previous work across all three datasets, we observe that AaltoDB shows the best performance (1.8\% EER), while HMOG DB (5.1\%) and HuMIdb (12.04\%) perform at a comparatively lower level. This can be attributed to the training dataset size difference, with Aalto DB having 60,000 viable users compared to only 69 and 328 for HMOG and HuMIdb, respectively. The lack of large-scale data is a common limitation when applying deep learning techniques for behavioural biometrics, and it also affects BehaveFormer. However, to overcome this limitation, we showcased how we could utilize the larger dataset to pretrain the weights of BehaveFormer before specializing it to a smaller dataset with transfer learning. This approach outperforms training from random weights, demonstrating the effectiveness of transfer learning in small dataset scenarios.

\section{Conclusion and Future Work} 
Our study introduces two novel components: STDAT, a transformer architecture capable of capturing both time and channel axis features, and BehaveFormer, a multi-modal architecture that combines two STDAT components for keystroke and IMU data. Our results demonstrate that BehaveFormer outperforms state-of-the-art keystroke CA systems in both keystroke-only and keystroke combined with IMU data scenarios. Notably, our observations indicate that incorporating all three types of IMU data (Accelerometer, Gyroscope, Magnetometer) significantly enhances the accuracy of keystroke biometrics.

Moreover, we explore the possibility of utilizing transfer learning to overcome the constraints of small datasets and learn a well-separated feature space. We believe that BehaveFormer can be applied to any behavioural biometric that involves multiple data modalities captured over time. In future work, we aim to extend the use of BehaveFormer and its transfer learning capabilities to other behavioural biometric applications.

{\small
\bibliographystyle{ieee}
\bibliography{main}

\begin{thebibliography}{10}\itemsep=-1pt

\bibitem{ACIEN2021104058}
A.~Acien, A.~Morales, J.~Fierrez, R.~Vera-Rodriguez, and O.~Delgado-Mohatar.
\newblock Becaptcha: Behavioral bot detection using touchscreen and mobile
  sensors benchmarked on humidb.
\newblock {\em Engineering Applications of Artificial Intelligence}, 98:104058,
  2021.

\bibitem{Acien2022}
A.~Acien, A.~Morales, J.~V. Monaco, R.~Vera-Rodriguez, and J.~Fierrez.
\newblock {TypeNet}: Deep learning keystroke biometrics.
\newblock {\em {IEEE} Transactions on Biometrics, Behavior, and Identity
  Science}, 4(1):57--70, 2022.

\bibitem{ba2016layer}
J.~L. Ba, J.~R. Kiros, and G.~E. Hinton.
\newblock Layer normalization, 2016.
\newblock cite arxiv:1607.06450.

\bibitem{hinton2008visualizing}
L.~Com and G.~Hinton.
\newblock Visualizing data using t-sne laurens van der maaten.
\newblock {\em Journal of Machine Learning Research}, 9:2579--2605, 2008.

\bibitem{electronics10141622}
L.~de~Marcos, J.-J. Mart{\'\i}nez-Herr{\'a}iz, J.~Junquera-S{\'a}nchez,
  C.~Cilleruelo, and C.~Pages-Ar{\'e}valo.
\newblock Comparing machine learning classifiers for continuous authentication
  on mobile devices by keystroke dynamics.
\newblock {\em Electronics}, 10(14):1622, 2021.

\bibitem{deb2019actions}
D.~Deb, A.~Ross, A.~K. Jain, K.~Prakah-Asante, and K.~V. Prasad.
\newblock Actions speak louder than (pass) words: Passive authentication of
  smartphone* users via deep temporal features.
\newblock In {\em 2019 international conference on biometrics (ICB)}, pages
  1--8. IEEE, 2019.

\bibitem{residualconnection}
K.~He, X.~Zhang, S.~Ren, and J.~Sun.
\newblock Deep residual learning for image recognition.
\newblock In {\em 2016 IEEE Conference on Computer Vision and Pattern
  Recognition (CVPR)}, pages 770--778, 2016.

\bibitem{hinton2012improving}
G.~E. Hinton, N.~Srivastava, A.~Krizhevsky, I.~Sutskever, and R.~R.
  Salakhutdinov.
\newblock Improving neural networks by preventing co-adaptation of feature
  detectors.
\newblock {\em CoRR}, abs/1207.0580, 2012.
\newblock cite arxiv:1207.0580.

\bibitem{ioffe2015batch}
S.~Ioffe and C.~Szegedy.
\newblock Batch normalization: Accelerating deep network training by reducing
  internal covariate shift.
\newblock In {\em Proceedings of the 32nd International Conference on
  International Conference on Machine Learning - Volume 37}, ICML'15, page
  448–456. JMLR.org, 2015.

\bibitem{adamopt}
D.~Kingma and J.~Ba.
\newblock Adam: A method for stochastic optimization.
\newblock {\em International Conference on Learning Representations}, 12 2014.

\bibitem{10.1145/3230820.3230829}
S.~Krishnamoorthy, L.~Rueda, S.~Saad, and H.~Elmiligi.
\newblock Identification of user behavioral biometrics for authentication using
  keystroke dynamics and machine learning.
\newblock In {\em Proceedings of the 2018 2nd International Conference on
  Biometric Engineering and Applications}, page 50–57, New York, {USA}, 2018.

\bibitem{Li_Cui_Wang_Zhang_Chen_Wu_2021}
B.~Li, W.~Cui, W.~Wang, L.~Zhang, Z.~Chen, and M.~Wu.
\newblock Two-stream convolution augmented transformer for human activity
  recognition.
\newblock {\em Proceedings of the AAAI Conference on Artificial Intelligence},
  35(1):286--293, May 2021.

\bibitem{10.1109/TC.2016.2622262}
A.~Mosenia, S.~Sur-Kolay, A.~Raghunathan, and N.~K. Jha.
\newblock {CABA}: Continuous authentication based on {BioAura}.
\newblock {\em IEEE Trans. Comput.}, 66(5):759–772, 2017.

\bibitem{10.1145/3338286.3340120}
K.~Palin, A.~M. Feit, S.~Kim, P.~O. Kristensson, and A.~Oulasvirta.
\newblock How do people type on mobile devices? observations from a study with
  37,000 volunteers.
\newblock In {\em Proceedings of the 21st International Conference on
  Human-Computer Interaction with Mobile Devices and Services}, pages 1--12,
  Taipei, Taiwan, 2019.

\bibitem{whitepaperpytorch}
A.~Paszke, S.~Gross, F.~Massa, A.~Lerer, J.~Bradbury, G.~Chanan, T.~Killeen,
  Z.~Lin, N.~Gimelshein, L.~Antiga, A.~Desmaison, A.~Kopf, E.~Yang, Z.~DeVito,
  M.~Raison, A.~Tejani, S.~Chilamkurthy, B.~Steiner, L.~Fang, J.~Bai, and
  S.~Chintala.
\newblock Pytorch: An imperative style, high-performance deep learning library.
\newblock In H.~Wallach, H.~Larochelle, A.~Beygelzimer, F.~d\textquotesingle
  Alch\'{e}-Buc, E.~Fox, and R.~Garnett, editors, {\em Advances in Neural
  Information Processing Systems}, volume~32. Curran Associates, Inc., 2019.

\bibitem{8698599}
S.~Rasnayaka and T.~Sim.
\newblock Who wants continuous authentication on mobile devices?
\newblock In {\em 2018 IEEE 9th International Conference on Biometrics Theory,
  Applications and Systems (BTAS)}, pages 1--9, 2018.

\bibitem{10145743}
D.~Senarath, S.~Tharinda, M.~Vishvajith, S.~Rasnayaka, S.~Wickramanayake, and
  D.~Meedeniya.
\newblock Re-evaluating keystroke dynamics for continuous authentication.
\newblock In {\em 2023 3rd International Conference on Advanced Research in
  Computing (ICARC)}, pages 202--207, 2023.

\bibitem{10.1109/TPAMI.2007.1010}
T.~Sim, S.~Zhang, R.~Janakiraman, and S.~Kumar.
\newblock Continuous verification using multimodal biometrics.
\newblock {\em IEEE Trans. Pattern Anal. Mach. Intell.}, 29(4):687–700, 2007.

\bibitem{7349202}
Z.~Sitová, J.~Šeděnka, Q.~Yang, G.~Peng, G.~Zhou, P.~Gasti, and K.~S.
  Balagani.
\newblock Hmog: New behavioral biometric features for continuous authentication
  of smartphone users.
\newblock {\em IEEE Transactions on Information Forensics and Security},
  11(5):877--892, 2016.

\bibitem{stragapede2022typeformer}
G.~Stragapede, P.~Delgado-Santos, R.~Tolosana, R.~Vera-Rodriguez, R.~Guest, and
  A.~Morales.
\newblock Typeformer: Transformers for mobile keystroke biometrics.
\newblock {\em arXiv preprint arXiv:2212.13075}, 2022.

\bibitem{stragapede_mobile_2022}
G.~Stragapede, P.~Delgado-Santos, R.~Tolosana, R.~Vera-Rodriguez, R.~Guest, and
  A.~Morales.
\newblock Mobile keystroke biometrics using transformers.
\newblock In {\em 2023 IEEE 17th International Conference on Automatic Face and
  Gesture Recognition (FG)}, page 1–6. IEEE Press, 2023.

\bibitem{https://doi.org/10.48550/arxiv.2206.02502}
G.~Stragapede, R.~Vera-Rodriguez, R.~Tolosana, and A.~Morales.
\newblock {BehavePassDB:} public database for mobile behavioral biometrics and
  benchmark evaluation.
\newblock {\em Pattern Recognition}, 134:109089, 2023.

\bibitem{Stragapede2022}
G.~Stragapede, R.~Vera-Rodriguez, R.~Tolosana, A.~Morales, A.~Acien, and G.~L.
  Lan.
\newblock Mobile behavioral biometrics for passive authentication.
\newblock {\em Pattern Recognition Letters}, 157:35--41, 2022.

\bibitem{stragapede2022mobile}
G.~Stragapede, R.~Vera-Rodriguez, R.~Tolosana, A.~Morales, A.~Acien, and
  G.~Le~Lan.
\newblock Mobile passive authentication through touchscreen and background
  sensor data.
\newblock In {\em 2022 International Workshop on Biometrics and Forensics
  (IWBF)}, pages 1--6. IEEE, 2022.

\bibitem{kbattacks}
F.~Towhidi, A.~A. Manaf, S.~M. Daud, and A.~H. Lashkari.
\newblock The knowledge based authentication attacks.
\newblock In {\em Proceedings of the International Conference on Security and
  Management (SAM)}, pages 1--5, Las Vegas, {USA}, 2011.

\bibitem{vaswani2017attention}
A.~Vaswani, N.~Shazeer, N.~Parmar, J.~Uszkoreit, L.~Jones, A.~N. Gomez,
  {\L}.~Kaiser, and I.~Polosukhin.
\newblock Attention is all you need.
\newblock {\em Advances in neural information processing systems}, 30, 2017.

\bibitem{1495927}
Q.~Xiao.
\newblock Security issues in biometric authentication.
\newblock In {\em Proceedings from the Sixth Annual IEEE SMC Information
  Assurance Workshop}, pages 8--13, New York, {USA}, 2005.

\bibitem{zhang2021understanding}
C.~Zhang, S.~Bengio, M.~Hardt, B.~Recht, and O.~Vinyals.
\newblock Understanding deep learning (still) requires rethinking
  generalization.
\newblock {\em Communications of the ACM}, 64(3):107--115, 2021.

\end{thebibliography}
}

\end{document}